# The Competing Effect of Initial Crack Depth Versus Chemical Strengthening Parameters on Apparent Fracture Toughness of Sodium Aluminosilicate Glass

Benedict Egboiyi [1,2], Trisha Sain [1],

[1] Department of Mechanical Engineering- Engineering Mechanics,

Michigan Technological University, Houghton, MI 49931, USA

[2] Science and Technology Division, Corning Incorporated, Corning, NY, USA

Email: boegboiy@mtu.edu (B. Egboiyi) and tsain@mtu.edu (T. Sain)

## Abstract

The widespread use of sodium aluminosilicate glass in many engineering applications due to its mechanical and optical properties (transparency, dielectric, etc.), has become common in recent years. However, glass, a brittle material, has its vulnerability to fracture. Processes such as heat treatment (heat tempering) or chemical strengthening through ion-exchange have been used to create residual stress profiles on the glass, in a bid to improve its fracture strength for applications such as in the automobile windshield, consumer electronics mobile communication devices display cover (e.g., smartphones and tablets), etc. However, failure still occurs, which is mostly catastrophic and expensive to repair. Therefore understanding, predicting, and eventually improving the resistance to damage or fracture of chemically strengthened glass is important to designing new glasses that would be tougher, while retaining their transparency. The relationship between the glass residual stress parameters such as the compressive stress (CS), depth of compression layer (DOL), and central tension (CT) versus apparent (effective) fracture toughness for different crack depth was investigated in this study using a Silicon Carbide particle blast plus ring-on-ring (RoR) test method, based on the IEC standard for retained biaxial flexural strength measurements. The particle blast plus RoR testing technique was determined as another experimental method that could be used to measure fracture toughness. The results also showed that improving the fracture resistance of glass via chemical strengthening requires a proper combination of CS, DOL, and CT, which is particularly dependent on the initial/existing crack (flaw) depth. It was determined that for a damage event involving the introduction of a shallow crack depth, the criterion for optimal resistance to fracture, in terms of apparent fracture toughness, is weighted more towards a high CS, than deep DOL while for a deep flaw damage event, it is more weighted towards deep DOL, than a high CS. These results provide a valuable piece of information in the design of a more robust glass in engineering applications.

## 1. Introduction

Glass is a widely used material in almost every branch of engineering due to its rigidity and transparency, despite its brittleness and varying mechanical strength. The last decades have seen the increasing use of glass in automobiles, architecture, consumer electronics, etc. While in an application, events such as impact, scratches, vibrations, etc. can result in undesirable or catastrophic failures. In architecture, the role of glass is gradually moving from non- structural members, such as windows and facades, to a load-bearing one, such as staircases, balconies, beams, etc. The presence of cracks in these structures could result in very costly failures and possible loss of lives. Theoretically, glass strength can be as high as 16 GPa due to its strong chemical bonds [1, 2, 3]. However, the presence of defects or cracks commonly referred to as "Griffith cracks" [4], introduced during manufacturing or handling, can drastically lower its strength by orders of magnitude. In comparison to plastic, glass is not only much harder and resistant to damage but also offers superior optical quality and a premium look, which is highly desirable in consumer electronics, automobile windshield, and interior design applications. Thus, the possibility of strengthening glass sheets allows the use of glass in applications, such as cover plates for mobile phones, tablets, notebooks, laptops, etc., [5, 6] where high strength and scratch resistance are desired due to everyday exposure. It has also been used in fabrication of the cockpit windows, high-speed train windshield, high end photochromic and white crown eyeglass lenses, and in the production of hard disk drives for computer data storage [7].

Several glass strengthening techniques exist in literature, as highlighted in [3]. These include fire polishing and etching to reduce the severity of the surface crack, surface coating with a polymer to minimize abrasion or scratches, crack pining and deflection to increase fracture toughness [8], introduction of surface compression through surface crystallization, glazing, thermal tempering, chemical strengthening, etc. [6].

Chemical strengthening (tempering) through ion-exchange has been widely used to improve the strength of glass [9, 10, 11, 12, 13]. This process first introduced by Kistler in 1962, is an effective method of strengthening thin glass. During this process, thin glass containing alkali ions, (such as Na+), is immersed in a bath of molten potassium nitrate ($KNO_3$) salt at temperature, $T_g$, high enough to overcome the activation energy for the glass inter-diffusion but lower than the strain point of the glass [14]. With time, due to chemical potential difference, sodium ions (Na+) migrate out of the glass, and their place is taken up by the relatively larger Potassium ions (K$^+$) on a one-to-one basis. The difference between the ionic radius of K$^+$ and Na$^+$ generates a residual compressive stress on the glass surface (Figure 1), which leads to the glass strengthening. The residual stress profile can be expressed as:

$$\sigma(x) = -\frac{BE}{1-\nu}(C(x) - \bar{C}) \tag{1}$$

where $\sigma(x)$ is the compressive stress at a depth 'x' measured from the top surface, $B$ is the linear network dilation coefficient or "Cooper coefficient", $E$ is the Young's modulus, $\nu$ is the Poisson's ratio, $C(x)$ is the concentration of the substituting ion at the depth $x$, and $\bar{C}$ is the average ion concentration in the bulk glass [15, 16], which must be subtracted from $C(x)$ to satisfy the force balance condition. Higher salt bath temperature increases the diffusion rate. i.e., the rate at which larger ions are being stuffed into the voids created by the smaller ions. At the same time, the high bath temperature induces visco-elastic stress relaxation and therefore reduces the build-up residual stress magnitude. It should be noted that eqn. (1) is only valid if no stress relaxation takes place, i.e. the ion-exchange temperature is well below the strain point of the glass [17, 18, 19, 20].

The knowledge of the residual stress field has a great practical interest as the strength of the glass can be linked to both the depth of the compressive layer (DOL) and the maximum surface compressive stress (CS) of glass articles after ion-exchange. The correlation between the glass residual stress parameters, such as compressive stress (CS), depth of compressive stress layer (DOL), central tension (CT), versus fracture toughness is not only a long-standing fundamental interest but also important for controlling the mechanical properties of chemically strengthened glass products [9, 21, 22]. It is to be noted that the CT represents the tensile stress distribution within the volume below the surface that is required to balance the CS.

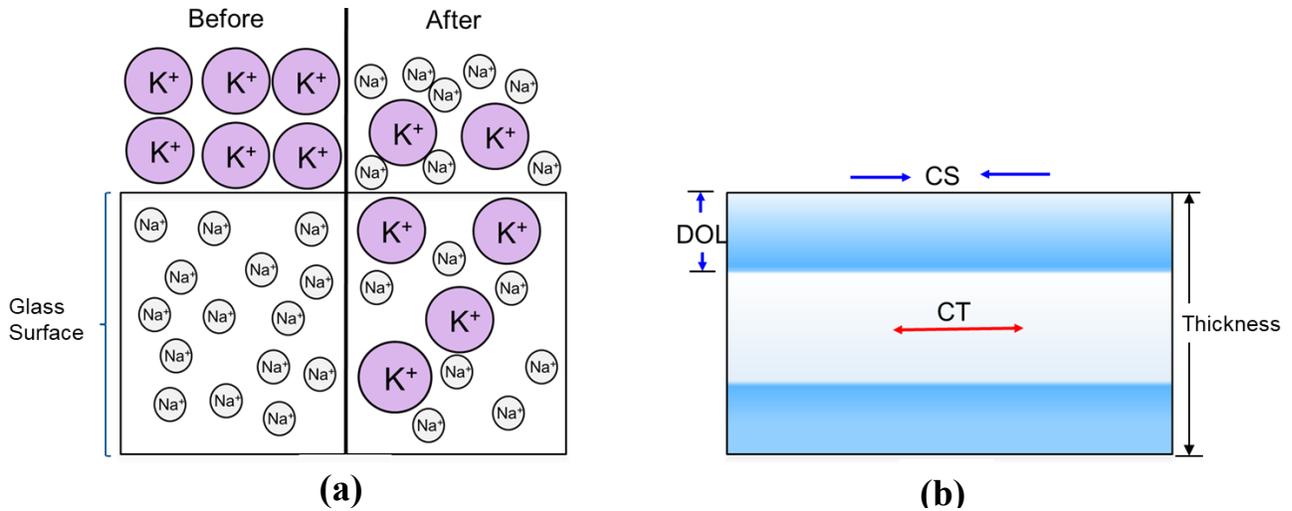

**Figure 1:** *Schematic of a typical ion-excahange process (a) replacement of smaller Na+ with larger K+; (b) formation of the compresson layer with a tensile stress region at the center of the glass.*

The presence of compressive stress is thought to improve the fracture strength of glass by creating residual stress fields around the tip of any pre-existing crack, thereby reducing the possibility of unstable crack growth [23, 24, 11, 25]. This effect can be explained using a concept of apparent fracture toughness [26, 27], in which the strengthening effect due to compressive residual stress is added to the inherent fracture toughness of the material. However, the relationship between the apparent fracture toughness and residual stress parameters (CS, DOL, CT) for a range of crack depths (length) is not substantially analyzed/studied with previously available experimental data [9, 28, 29, 30, 31].

Based on the current state of the art, in this work, we demonstrate using ion-exchanged glass samples with various combinations of CS, DOL, and CT, how cracks at different depths (lengths), influence the fracture strength of an ion-exchanged (chemically strengthened) glass. We also show how the "apparent" critical stress intensity factor (toughness) of a chemically strengthened sodium aluminosilicate (SAS) glass changes with respect to its residual stress parameters, for various crack depths comparing experimental measurements and the analytical model predictions.

2. **Experimental procedure**

The samples used in the experiment were 0.8mm thick, 50 mm x 50 mm coupons sodium aluminosilicate glass (SAS) made by Corning® Incorporated, with properties and nominal composition referenced in [32, 33]. These samples were ion-exchanged in an electric furnace. Prior to ion-exchange, all the specimens were preheated to a temperature 250°C for 15 minutes to remove any residual stress in the glass and to prevent thermal shock of the glass when placed in the ion-exchange bath. The glasses were placed in an ion-exchange bath at different salt ($KNO_3$) concentrations, bath conditions, and durations. At the end of the ion-exchange cycle, the glass specimens were thoroughly rinsed and cleaned with deionized water. The specimens were labeled as A, B, C, D, E, F, and G as per the ion–exchange condition. The magnitude of the CS and DOL were measured by using the Iterated Wentzel-Kramers-Brillouin (IWKB) system [34] and plotted as shown in Figure 2. The (approximate) magnitude of the CT was calculated from the measured CS, DOL and the sample thickness, $t$, using the equation as given below,

$$CT = (CS * DOL) / (t - 2 * DOL). \qquad (2)$$

Eqn. (2) had been derived from force balance condition and assuming a linear stress profile (as opposed to the actual non-linear nature of the stress profiles (Figure 2)). It is to note that the specimen H is a non-chemically strengthened (as-received) glass of the same glass composition (see Table 1). The magnitude of the CS, DOL, CT (calculated), and bath conditions for the test specimens are summarized in Table 1.

After the ion-exchange and cleaning process, the specimens were taped on one side using 3M™ 471Vinyl adhesive tape before the abrasion procedure. This is necessary to keep the glass pieces together upon fracturing for fractographic analysis. The non-taped side of the specimen was subjected to air-forced silicon carbide abrasion at different pressures. The abrasion pressures used in this experiment were 1, 3, 8, and 15 psi using 90 grit silicon carbide (SiC) particles, respectively. These pressures were selected to create different levels of initial crack depth. The abrasion time per sample was 5 seconds. The abraded area was approximately 6mm in diameter and located at the center of the specimens. The 90 grit SiC particle were sieved prior to use to minimize variation in the data. After abrasion, the samples were sat in the test lab (temperature, T = 22°C ± 2°C and relative humidity, Rh = 50 % ± 5%) for a minimum of 24hrs, to eliminate any possibility of delayed/unstable crack growth, prior to fracture strength testing [35].

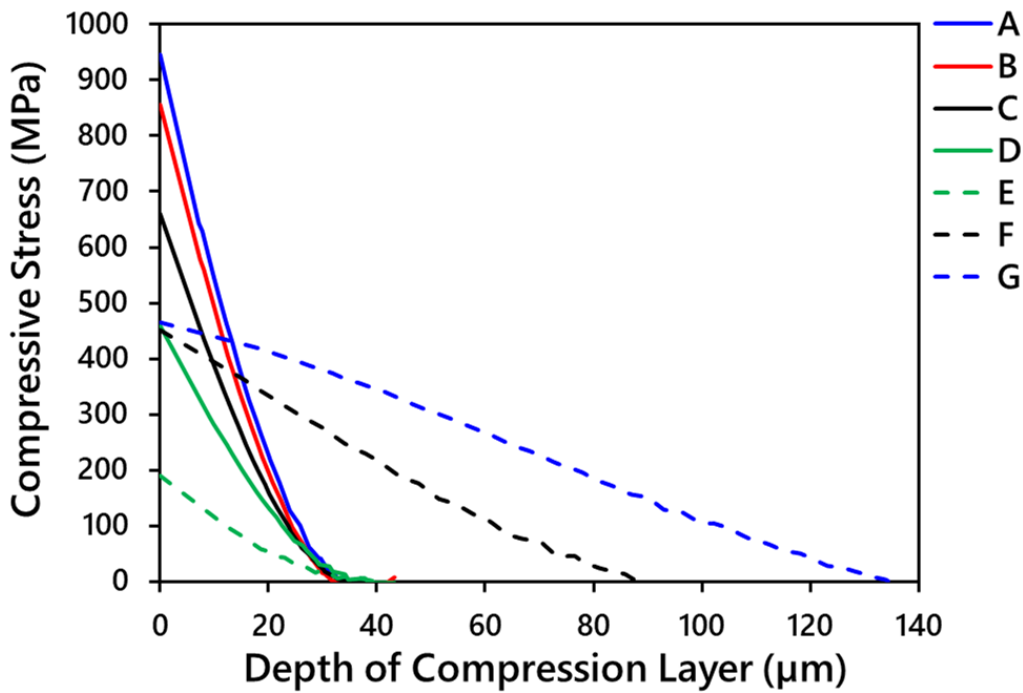

*Figure 2:* *Residual compressive stress profiles for the various Ion-exchanged specimens.*

After 24hrs post abrasion, the specimens were placed in a ring-on-ring (RoR) setup, having a load ring diameter of 12.7mm, support ring diameter of 25.4mm, and a contact radius of 1.6mm, mounted in a screw-driven machine (Instron). The specimens were subsequently loaded to failure at a constant displacement rate of 1.2mm/min, such that the abraded area experiences tensile stress (faces the support ring). A Teflon tape was placed between the specimen non-taped face and the support ring to prevent friction damage from direct contact with the hardened stainless-steel the ring is made of (Figures 3a, b). Five specimens were tested per abrasion pressure for the various ion-exchange condition. Upon fracturing of the glass, fractography analysis was conducted to measure the width and depth of the crack for the various abrasion pressures using a Nikon Eclipse LV100 compound optical microscope equipped with NIS Elements measurement software. All measurements were done using the same magnification and lighting conditions to ensure comparable measurements between the samples.

.

Table 1: Ion-exchange parameters of chemically strengthened glass

| Specimen | Bath salt concentration | Ion-Exchange Time (hr) | Bath Temperature (°C) | CS (MPa) | CT (MPa) | DOL (μm) |
|---|---|---|---|---|---|---|
| A | 95% KNO$_3$ and 5% K$_2$CO$_3$ | 2.8 | 390 | 946 | 44.5 | 34.4 |
| B | 99% KNO$_3$ and 1% NaNO$_3$ | 2.8 | 390 | 856 | 38.0 | 32.6 |
| C | 90% KNO$_3$ and 10% NaNO$_3$ | 3.0 | 390 | 660 | 31.1 | 34.5 |
| D | 70% KNO$_3$ and 30% NaNO$_3$ | 4.0 | 390 | 462 | 24.5 | 38.3 |
| E | 25% KNO$_3$ and 75% NaNO$_3$ | 5.2 | 390 | 190 | 10.0 | 38.1 |
| F | 79% KNO$_3$ and 21% NaNO$_3$ | 9 | 450 | 463 | 65.3 | 88 |
| G | 93% KNO$_3$ and 7% NaNO$_3$ | 28 | 450 | 465 | 117.5 | 134.3 |
| H | - | - | - | - | - | - |

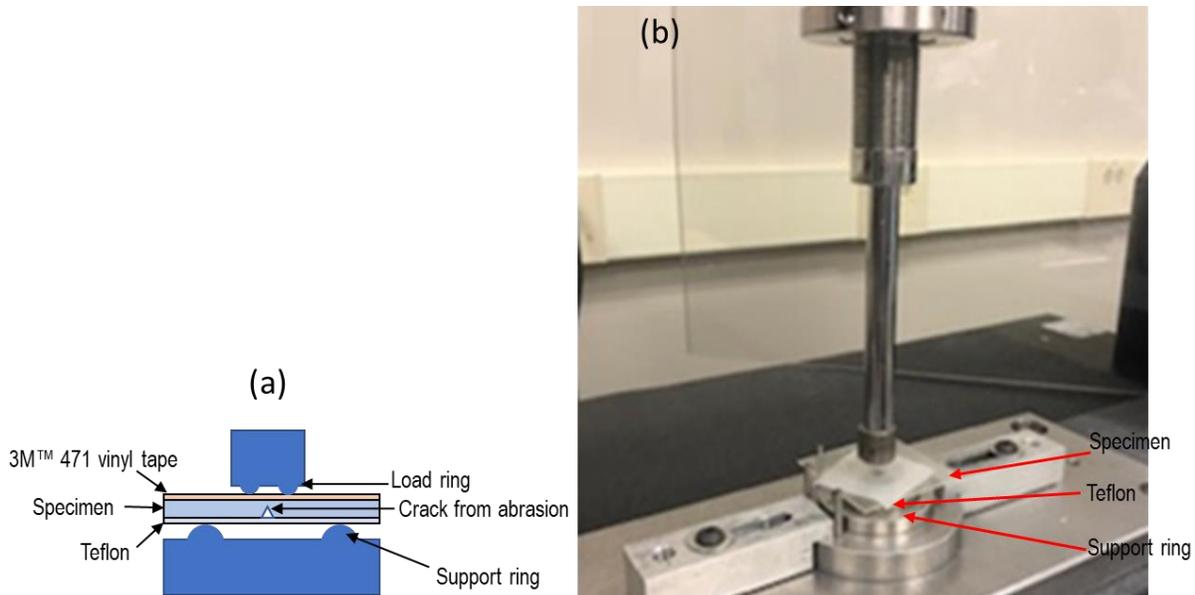

*Figure 3: Ring on Ring test setup (a) schematic section view illustrating test specimen setup (b) test specimen subjected to loading*

## 2.1. Computation of the fracture stress from the experimentally measured fracture load

In this present study the maximum applied tensile stress (fracture stress) $\sigma_f$ for all the specimens A -H is computed from the experimentally measured fracture load using the equation, as given below (based on the small deflection theory [36, 37, 38]),

$$\sigma_f = \frac{3P}{2\pi t^2}\left[(1+v)\ln{R_s}/{R_L} + (1-v)\frac{(R_s^2-R_L^2)}{2R^2}\right], \quad (3)$$

where $P$ is the fracture load, $R_L$ is the load ring radius, $R_s$ the support ring radius, $v$ is the Poisson's ratio, and R is the so-called equivalent radius of the square sample given as $R = 0.27(l_1 + l_2)$; where $l_1$, $l_2$ are the length and width of the specimen. The computed fracture stresses obtained were significantly high for the 1psi and 3psi abraded samples, due to the effect of membrane stresses, hence, a non-linear stress analysis was conducted using large deflection based FEA technique following earlier study [9] to correct for this (see Figure 4). The obtained large deflection-based fractures stresses, $\sigma_f$ were used in subsequent analysis. The important findings related to the experimental data are described in detail in the following sections.

It is important to note that the glass strength evaluation is strongly dependent on the existing conditions of the glass surface (even for a chemically strengthened glass). The fracture strength or fracture stress, $\sigma_f$, for the non-chemically strengthened glass, H, can be computed using the classical Griffith [39] equation as,

$$\sigma_f = \frac{1}{Y}\sqrt{\frac{2E\gamma}{c}} = \frac{K_{Ic}}{Y\sqrt{c}} \quad (4)$$

where, $E$ is the Young's Modulus, $K_{Ic}$ is the fracture toughness (i.e., the critical stress intensity factor in mode I fracture), $c$ is the crack depth (the critical crack depth or the check depth), Y is the shape factor, a dimensionless constant (containing all geometric factors including $\pi^{0.5}$) related to the crack geometry and $\gamma$ is the fracture surface energy of the glass. However, for the chemically strengthened glass (A - G) a residual compressive stress, $|\sigma_{rcs}|$, is present around the crack, which reduces the intensity of the stress experienced by the crack. Note that $|\square|$ represents the integral of the residual compressive stress (resultant) profile distribution along the crack depth. The applied fracture stress (tensile) must overcome the residual compressive stress before the crack experiences tensile stress [5, 22]. Thus, the applied fracture stress is given as:

$$\sigma_f = \frac{K_{Ic}}{Y\sqrt{c}} + |\sigma_{rcs}| \tag{5}$$

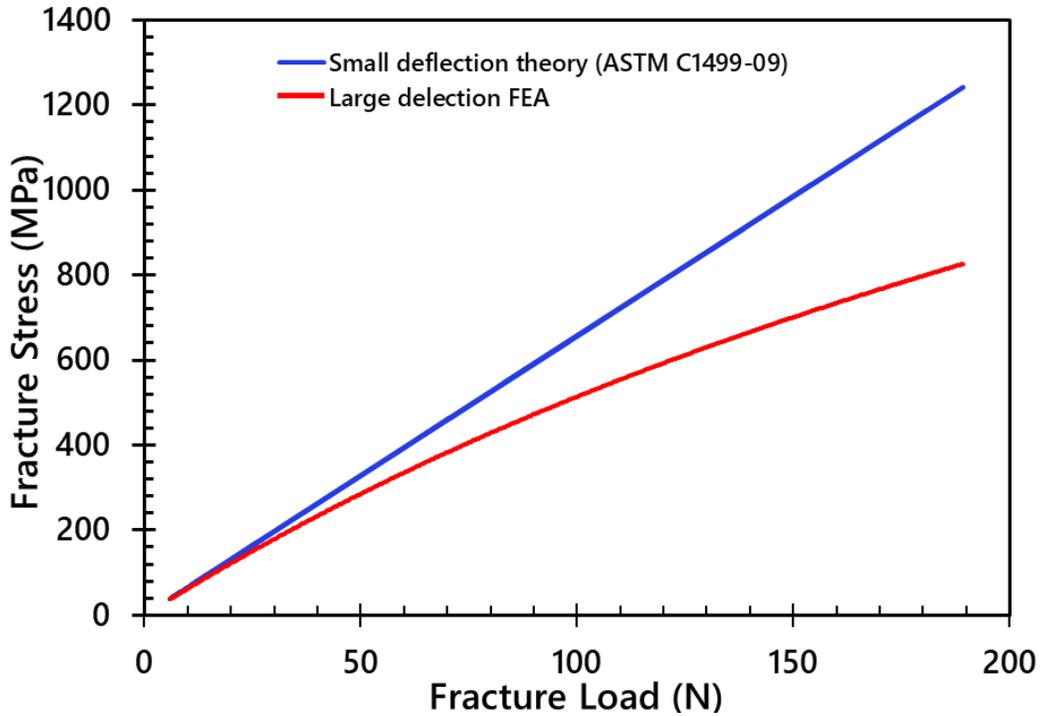

*Figure 4: Ring-on-Ring stress as a function of the experimentally measured fracture load: small deflection (ASTM) and large deflection (FEA) deflection solutions*

2.2. Computing apparent fracture toughness from the fracture stress

The experimentally measured apparent fracture toughness, $K_{IC}^a$, for the specimens A-H is calculated using the measured fracture stress and the measured critical crack depth [39] using:

$$K_{IC}^a = \sigma_f Y \sqrt{c} \tag{6}$$

The values of the shape factor *Y* used in the calculation was determined from the fractographic analysis measured width and depth of the crack following ASTM c1322-05b and Quinn 2016 [36, 40] for all the specimens tested.

It should be noted that for an existing surface crack under the applied tensile stress normal to the crack face, and a residual compressive stress distribution along the crack, the stress intensity factor $K_I$, (which is a measure of the intensification of the applied stress caused by the presence of the crack) is given by:

$$K_I = K_I^a + K_r, \tag{7}$$

where $K_r$ is the stress intensity factor associated with the residual stress field from the chemical strengthening process [22, 41, 42] and $K_I^a$ is the apparent stress intensity factor due to the applied load. $K_I^a$ is the critical value that quantifies the damage or fracture resistance of chemically strengthened glass. The summation of the stress intensities can be written based on the principle of superposition, given that both stresses are causing the same mode of fracture (i.e. mode 1 fracture loading condition). Following the condition of unstable crack growth, $K_I = K_{IC}$, eqn. (7) can be re-written as,

$$K_{IC}^a = K_{IC} - K_r \tag{8}$$

For a compressive residual stress, $K_r$ is negative ($K_r < 0$) resulting in an apparent toughening of the glass. In the presence of no residual compressive stress (as in specimen H) $K_r = 0$. Therefore $K_{IC}^a = K_{IC}$.

## 3. Analytical procedure

Two analytical based approaches, the Green's function and Crack tip residual stress analytical approaches for determining $K_{IC}^a$ are explored in this study and results obtained are compared with experimental data.

### 3.1. Computing apparent fracture toughness from Green function approach.

Assuming a linear profile for the compressive stress, with a surface compressive stress CS linearly decreasing to zero at the depth DOL (see Figure 5), the residual stress distribution profile may be represented by a first order linear expression [22, 26, 43], given as;

$$\sigma_{rcs}(x) = -CS\left(1 - \frac{x}{DOL}\right) \text{ if } x \leq DOL \tag{9}$$

$$\sigma_{rcs}(x) = CT \text{ if } x > DOL,$$

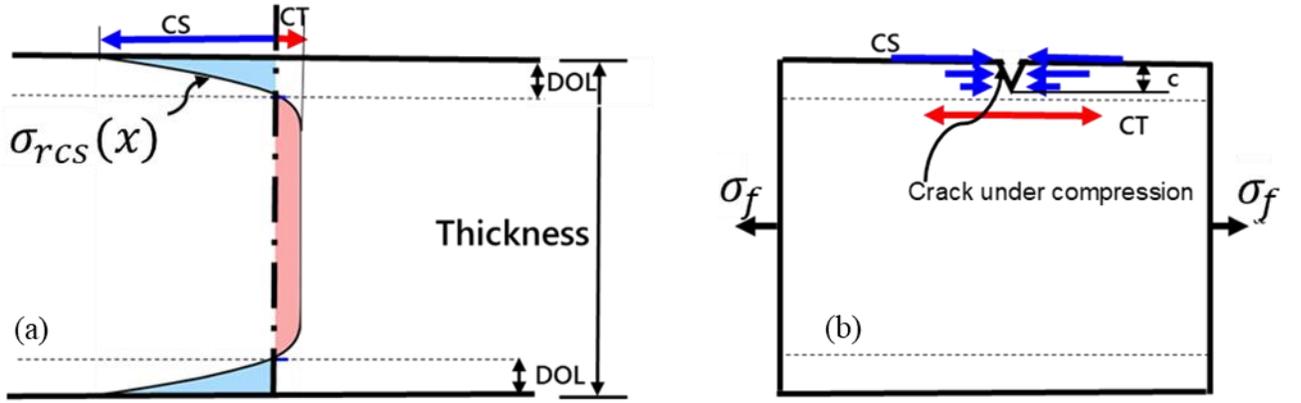

**Figure 5:** *Schematic representation of the residual stress distribution (a) across the specimen the thickness (b) along a crack of depth c for a chemically strengthened glass*

The residual stress intensity factor, $K_r$ can be computed from the given residual stress field $\sigma_{rcs}(x)$ by integrating using the Green's function approach [41, 43] as;

$$K_r = \frac{2Y\sqrt{c}}{\pi} \int_0^c \frac{\sigma_{rcs}(x)}{\sqrt{(c^2 - x^2)}} dx \tag{10}$$

Upon substituting eqn. (9) in eqn. (10) and integrating across the crack depth c, it simplifies to:

$$K_r = -YCS\sqrt{c}\left(1 - \frac{2c}{\pi DOL}\right), \text{ if } c \leq DOL \tag{11}$$

Also, if $c \geq DOL$, eqn. (10) takes the form as;

$$K_r = Y\sqrt{c}\left(\frac{\pi}{2}(CS + CT)\sin^{-1}\left(\frac{DOL}{c}\right) - \frac{2cCS}{\pi DOL} + \frac{2CS}{\pi}\sqrt{\left(\left(\frac{c}{DOL}\right)^2 - 1\right)} - CT\right) \tag{12}$$

Substituting eqn. (11) and eqn. (12) into eqn. (8), gives the analytically computed $K_{IC}^a$, as:

$$K_{IC}^a = K_{IC} + \left|YCS\sqrt{c}\left(1 - \frac{2c}{\pi DOL}\right)\right|, \text{ if } c \leq DOL \tag{13}$$

$$K_{IC}^a = K_{IC} + Y\sqrt{c}\left(\frac{\pi}{2}(CS + CT)\sin^{-1}\left(\frac{DOL}{c}\right) - \frac{2cCS}{\pi DOL} + \right.$$
$$\left. \frac{2CS}{\pi}\sqrt{\left(\left(\frac{c}{DOL}\right)^2 - 1\right)} - CT\right), \quad if\ c \geq DOL \quad (14)$$

Note that eqn. (13) and eqn. (14) simplify to $K_{IC}^a = K_{IC}$, in the absence of residual stress (no chemical strengthening).

The Green's function based approach originally considers the hypothesis of the crack geometry conservation (both shape and depth) during ion exchange [22, 41, 43]. However, since the abrasion (damage) was done on the samples post chemical strengthening in this study, the experimentally measured crack depth for the individual samples was used in the analytical computation.

3.1. Computing apparent fracture toughness from crack tip residual stress approach.

While the Green's function-based approach considers the compressive stress distribution along the entire crack depth to calculate the stress intensity factor $K_r$, the crack tip residual stress approach involves $K_r$ computation considering the residual stress contribution only at the crack tip. In Figure 6 the residual compressive stress at the crack tip is plotted for the various ion-exchange conditions considered in this study. It is shown to vary in magnitude as a function of two different crack depths (c=10μm and c=40μm). The compressive residual stress at the crack tip for the specimen A-G, are denoted as $\sigma_{rcs\_*}$ where "*" in the subscript denotes the sample alphabetic name. If a crack depth of c = 10μm is assumed, we can roughly estimate from the given stress profile that the compressive stress at the tip follows the order as: $\sigma_{rcs\_A} > \sigma_{rcs\_B} > \sigma_{rcs\_G} > \sigma_{rcs\_F} > \sigma_{rcs\_C} > \sigma_{rcs\_D} > \sigma_{rcs\_E} > 0$. As the crack depth goes deeper to c = 40μm, the stress at the crack tip becomes $\sigma_{rcs\_A} \approx \sigma_{rcs\_B} \approx \sigma_{rcs\_C} \approx \sigma_{rcs\_D} \approx \sigma_{rcs\_E} \approx 0$ with only exception for F and G as, $\sigma_{rcs\_G} > \sigma_{rcs\_F} > 0$.

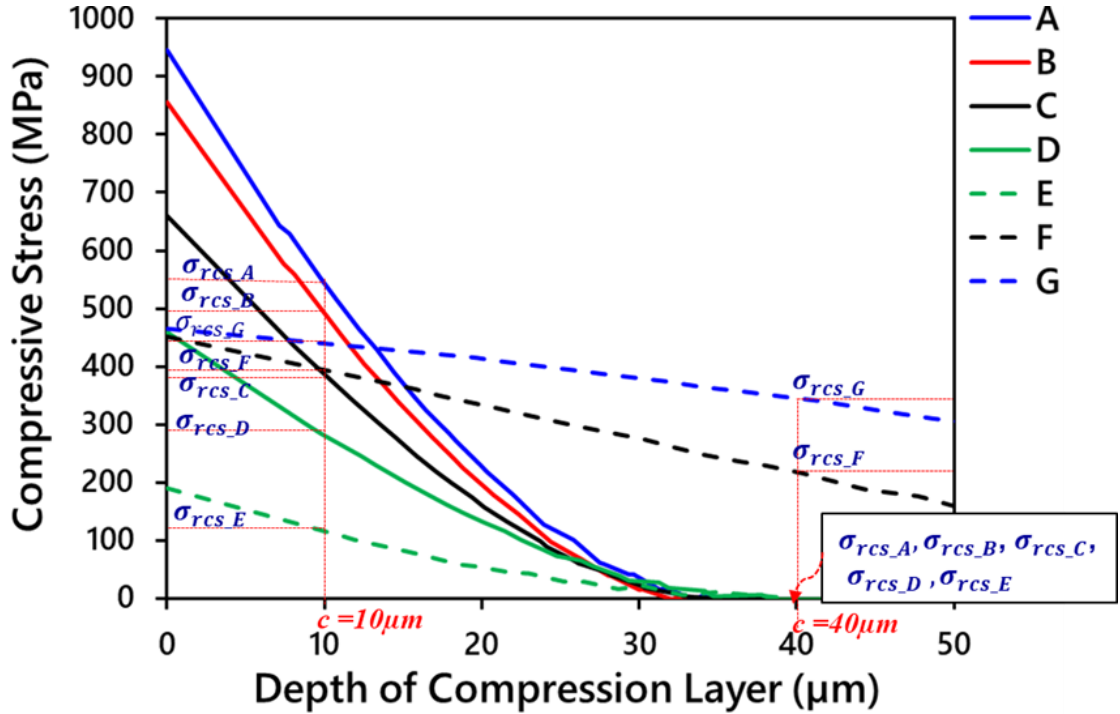

*Figure 6:* Stress profile of samples showing the magnitude of the residual comptressive stress at a crack depth c=10μm and c = 40μm.

The apparent fracture toughness for crack tip residual stress method is then calculated using:

$$K_{IC}^a = K_{IC} + \sigma_{rcs\_*} Y\sqrt{c}, \tag{15}$$

where $\sigma_{rcs\_*}$ is the residual compressive stress at the crack tip of the individual specimen A-G measured directly from the stress-profiles as shown in Figure 6.

## 4. Results and discussion

4.1. Crack depth

Fractograpghic analysis of all the tested dspecimens revealed that the fracture origin (i.e., the point of crack initiation in glass fracture) for all the specimens fractured in the RoR occurred from the abraded spot as illustrated in Figures 7a,b. The crack depth (i.e., crack depth/length at which unstable fracture occurs) for each fractured specimen was determined. It should be noted here that a crack in a glass specimen can be thought of as a generally (shallow) fracture having an arrested crack front as shown in Figure 8. The crack front extends into the material from the surface to a depth that can be measured, defined by the arrest mark feature that extends furthest from the surface (and is less than the thickness of the specimen) [36, 40, 44]. A few examples of the crack geometry from this experiment are shown in Figure 9.

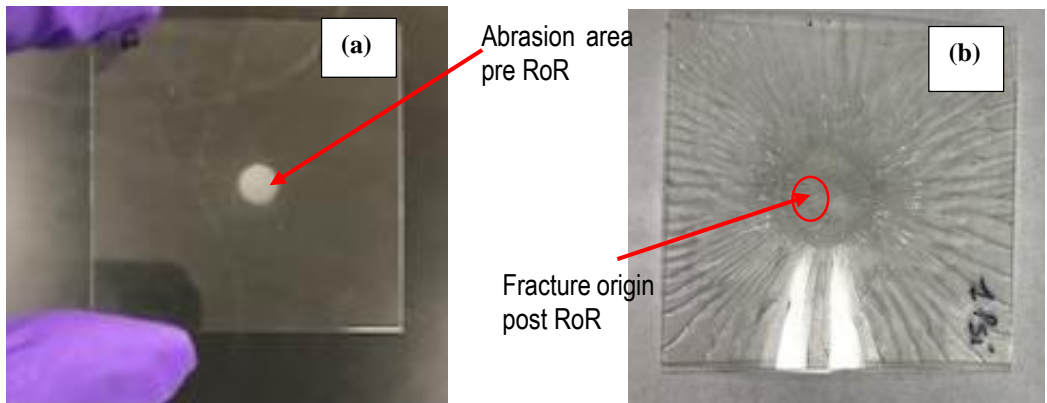

*Figure 7: Comparison of the damage surface before and after fracture for a 1psi abrasion pressure condition: (a) after abrasion (b) after fracturing via RoR.*

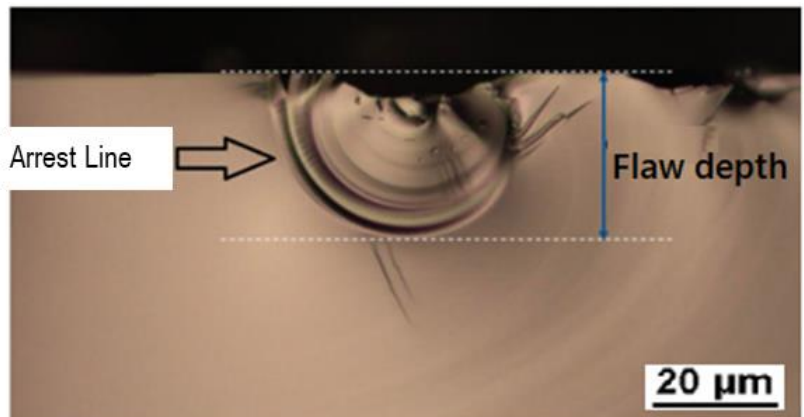

*Figure 8: Example of arrest line which forms the crack depth.*

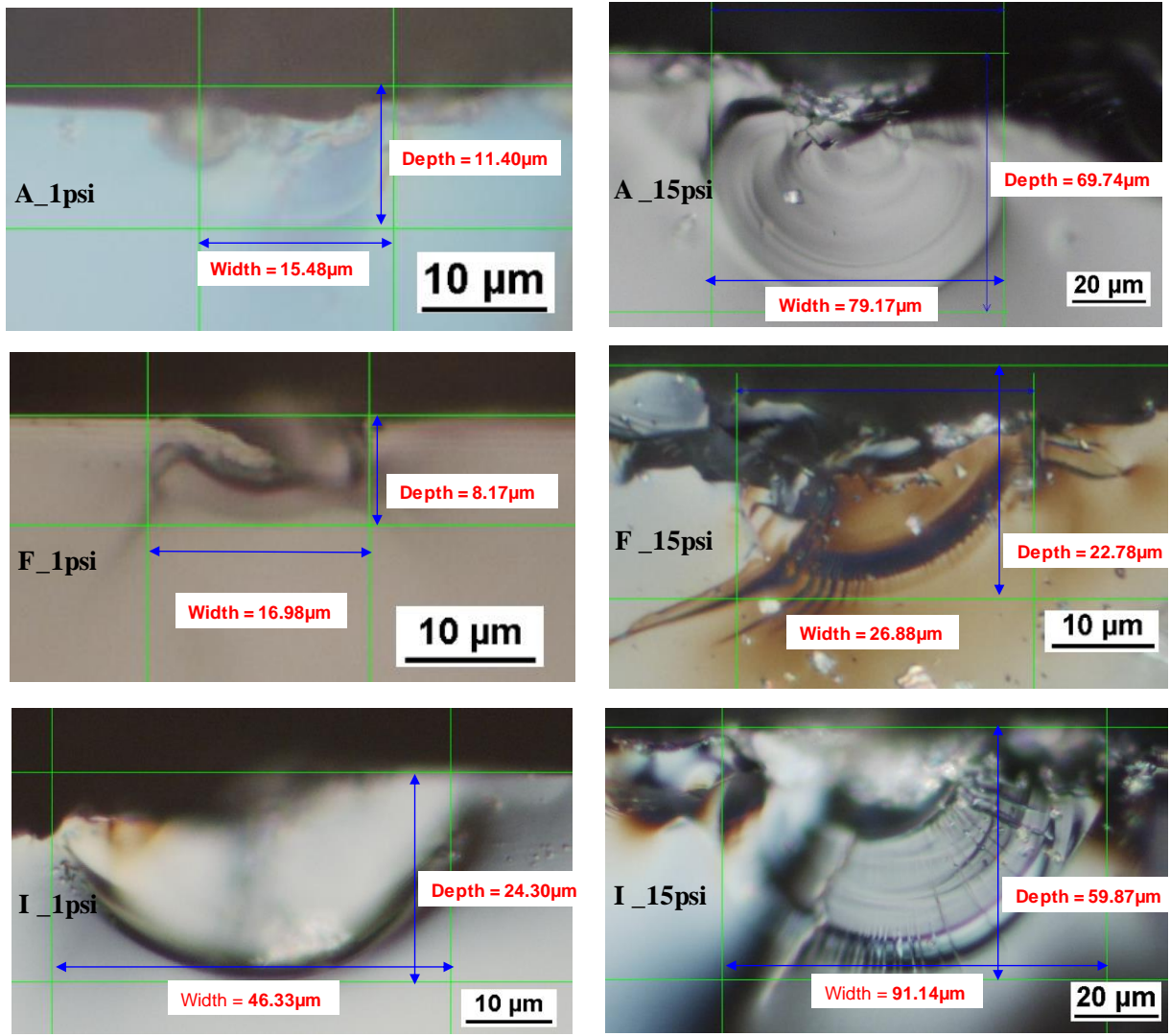

*Figure 9: Examples of crack shape and crack depth for specimens A, F, and I, fractured in ROR after damage introduction at 1, and 15psi abrasion pressure, respectively.*

4.1.1. Crack depth as a function of abrasion pressure

Figure 10 shows the measured crack depth as a function of abrasion pressure. It shows that for the samples A-H, as the abrasion pressure increases, the deeper the crack penetrates. Hence an increase in abrasion pressure leads to an increase in the crack depth, as expected. It is observed that the average critical crack depth for H (non- chemically strengthened sample) is higher than the chemically strengthened glasses, at both 1psi and 3psi. This is due to the absence of residual compressive stresses in the non-chemically strengthened glass. As the abrasion pressure increases beyond 3psi, the crack depths for specimens A, B (with high surface CS), C, D, E, and H, increase quite significantly but less dominantly for F and G (with deep DOL). This is due to the presence of deeper layer of compressive stresses in F and G, that helps to mitigate the effect of the applied tensile stress causing crack propagation. Higher abrasion pressure (greater than 15psi) may be required to generate the similar depth of cracks as those in specimen A-E, and H. It is to note that the 1psi abrasion pressure is observed to create crack depths to those generated from glass handling or machining processes [6] while the 15psi abrasion pressure creates crack depth within the domain of those introduced through sharp contact damage, such as drop of consumer electronics mobile commination hand held devices cover glass on rough surfaces [30, 45].

4.1.2. Crack depth as a function of surface CS and DOL per thickness

The variation of crack depth as a function of surface CS was studied for cracks created by the 1psi and 15psi abrasion pressure as shown in Figure 11(a). The 1psi abrasion is observed to create a crack depth c < 20μm which we categorize shallow crack while the 15psi creates a crack depth c > 20μm, categorized also in this study as deep cracks. It is observed that as the surface CS increases the crack depth decreases for the 1psi abrasion pressure. However, for the 15psi abrasion pressure the crack depth remains approximately constant as the surface CS increases. These observations suggest that high surface CS helps to mitigate against the creation of shallow cracks resulting from everyday handling of the glass, surface polishing etc. but not so much for deeper cracks that are typically created by sharp contact damage.

The variation of the crack depth as function of DOL per thickness was also investigated for the 1psi and 15psi abrasion pressures as shown Figure 11 (b). It is observed that as the DOL increases the crack depth decreases for both the 1psi and 15psi abrasion pressures (more significantly for the 15psi damage). In contrast to the case of increasing surface CS, the crack depth is significantly decreased by increasing DOL for deeper cracks. This suggest that the presence of deep DOL helps to mitigate against the introduction (or penetration) of deeper or severe cracks.

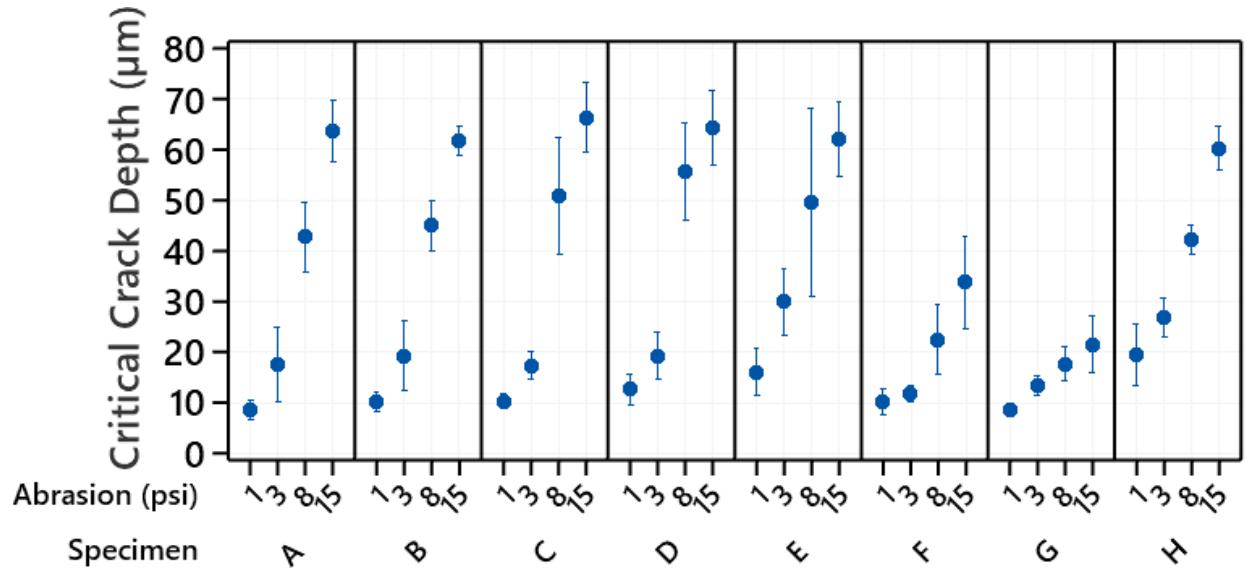

*Figure 10: Average crack depth as a function of abrasion pressure for all the samples.*

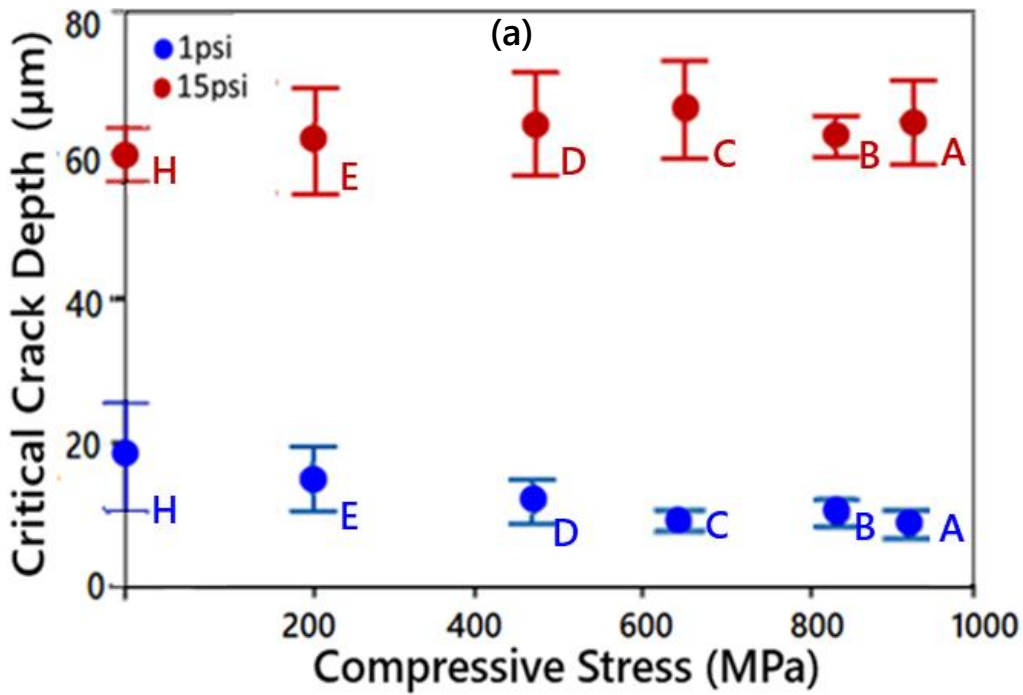

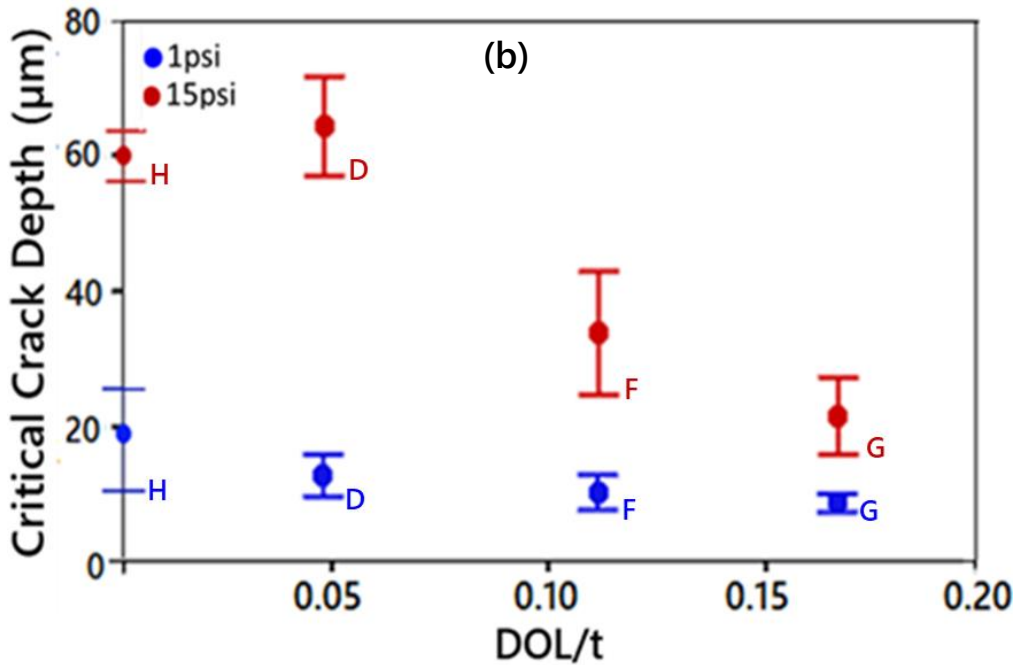

***Figure 11:*** *Crack depth for 1psi and 15psi abrassion as a function of (a)surface CS at constant DOL (35 ± 3μm) (b) DOL per thickness at constant surface CS (463 ± 4MPa)*

4.2. Fracture stress as a function of abrasion pressure

Figure 12 shows the computed fracture stress as a function of the abrasion pressures. The fracture stress $\sigma_f$ in RoR, for the non-chemically strengthened glass, H (CS = 0MPa, DOL = 0μm) is found to be the lowest compared to those of the chemically strengthened glass (for the same glass composition). This observation basically reflects the benefits of the residual compresive stress being present in the case of the chemically strengthened glass.

For the chemically strengthened glasses, A-G, the applied fracture stress is observed to decrease as the abrassion pressure increases. This is because as the abrasion pressure increased, the resulting crack depths also increase. Sample A (CS =946MPa, DOL = 34.4μm) has the highest fracture stress at 1 and 3psi abrasion pressures. The most likely reason for this observation is that the depth of cracks generated by lower abrasion pressures are still well within the DOL (c < DOL), and the fracture resistance proportionally increases as a function of the surface residual compressive stress in that case. For c < DOL, the crack is contained within the compressive stress regime and the entire crack depth is under the influence of the residual compressive stress distribution (Figure 5). Hence, the fracture stress increases as a function of the magnitude of the integral residual compressive stress, $|\sigma_{rcs}|$, along the crack depth, which is proportional to the surface CS. As the abrasion pressure increases beyond 3psi, the crack depths generated go deeper or often exceed the DOL (i.e c ≥ DOL). Hence, the effect of the compressive residual stress reduces along the crack depth and vanishes at the crack tip. Thus, the fracture strength of A ( and B, C, D, E ) with shallower DOL drops significantly for higher abrasion pressures. Conversely, the fracture stress of G (CS = 465MPa, DOL =134.5µm) is found to be significantly higher at even 8psi and 15psi abrasion pressures, than the rest of the specimens. This is because the crack depths generated at these abrasion presures are still well within the DOL of G (i.e c < DOL), which is considerably higher than the DOL of the other specimens.

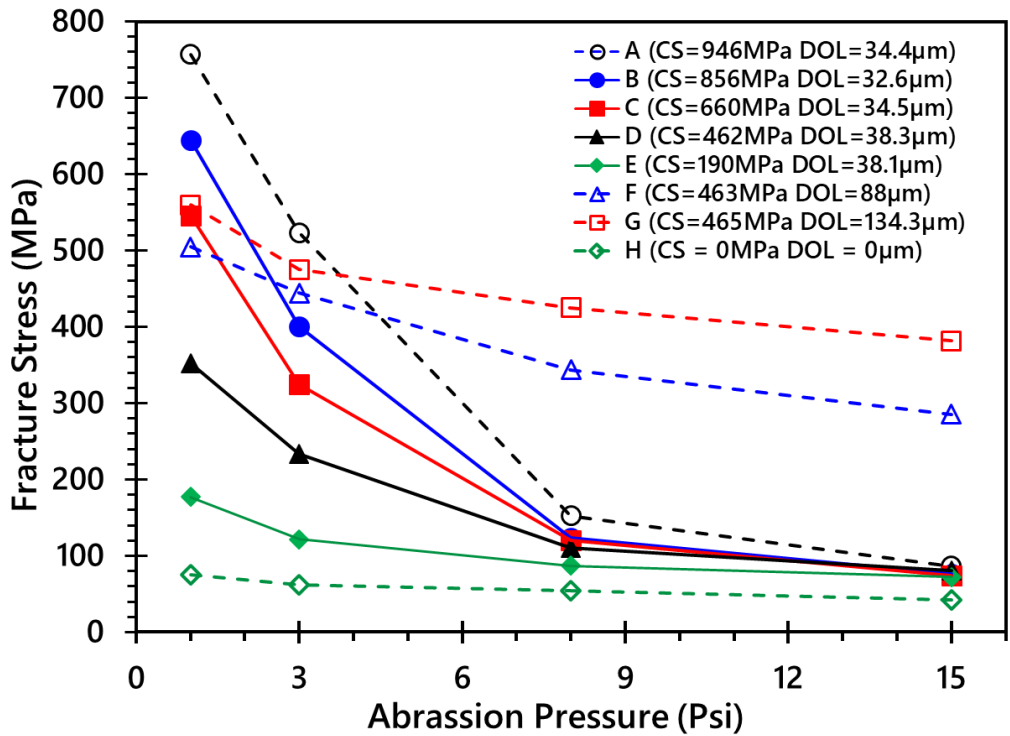

**Figure 12:** *Fracture stress as a fuction of abrasion pressure.*

4.3. Effect of CS - DOL combination on the fracture stress of ion-exchanged samples for shallow

and deep cracks

Figure 13 shows the measured fracture stresses for the samples A-H, as a function of average critical crack depths created by the 1, 3, 8, and 15psi abrasion pressures, respectively. The fracture stress is observed to decrease with increase in crack depth. The fracture stress of the non-chemically strengthened glass H, is observed to be significantly lower than those of the chemically strengthened glass across the crack depth. This is due to the fact that, for the non-chemically strengthened glass, there is no compressive residual stress present along the crack depth. i.e. $|\sigma_{rcs\_H}| \approx 0$. The higher magnitude of the fracture stresses for the chemically strengthened glasses A-G, is due to the presence of compressive residual stress along the entire crack depth (i.e. $|\sigma_{rcs}| > 0$), which ( in addition to the inherent strength of the glass) must be overcome, before fracture can occur.

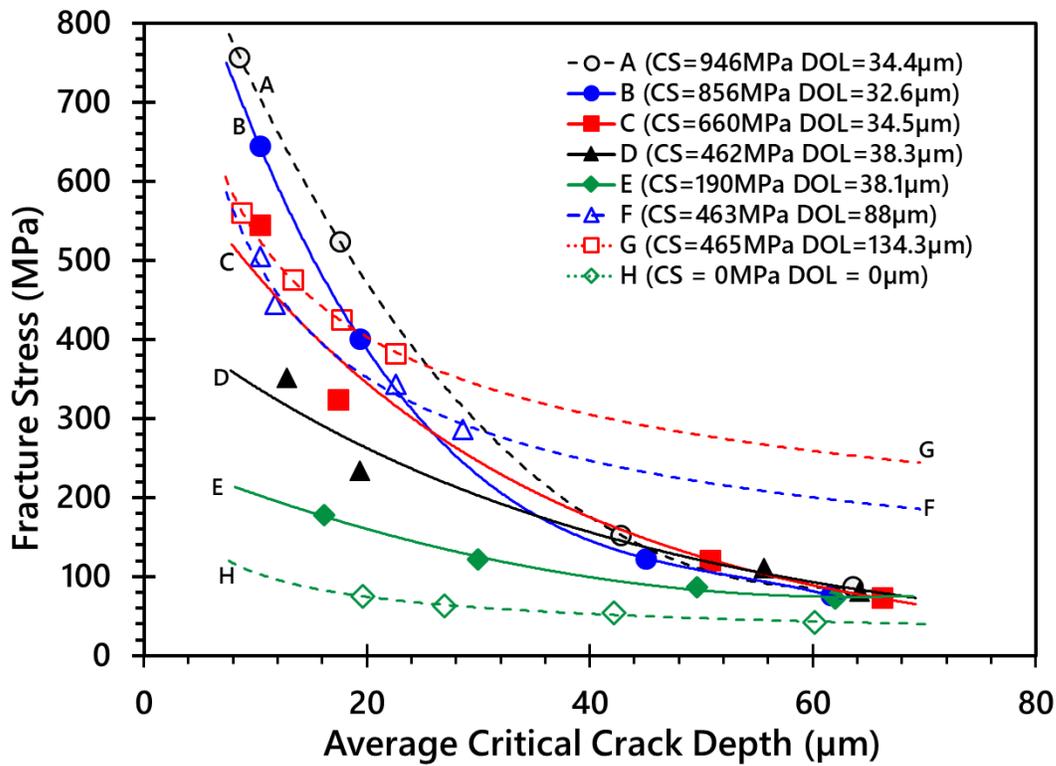

*Figure 13:* *Fracture stress as a fuction of critical crack depth.*

Within the shallow crack depth domain (c < 20μm created by 1psi abrasion pressure), for specimens A – E having a approximately constant DOL and changing surface CS, the measured fracture stress is observed to increase with increasing surface CS. It is significantly higher for A followed by B, C, D, and E. This is because the magnitude of $|\sigma_{rcs\_A}| > |\sigma_{rcs\_B}| > |\sigma_{rcs\_C}| > |\sigma_{rcs\_D}| > |\sigma_{rcs\_E}|$. For samples D, F, and G having an approximately constant surface CS and changing DOL, the fracture stress is observed to increase with increase in DOL. Also in Figure 13, if we consider a fixed crack of depth, c = 10μm within the shallow crack domain ( c < 20μm), and compare specimen A which has a high CS with specimen G which has a much deeper DOL but lower surface CS than A, the fractures stress of A is observed to be significantly higher than that of G. This is because for the crack depth c = 10μm, the magnitude of $|\sigma_{rcs\_A}| > |\sigma_{rcs\_D}|$.

For a deeper crack depth domain (c > 20μm created by 15psi abrasion pressure), the fracture stresses for specimens A-E decreases significantly, due to the reduced or almost vanishing magnitude of $|\sigma_{rcs}|$, while for specimens D, F, and G having an approximately constant surface CS and changing DOL, the fracture stress is observed to increase with increase in DOL. The fracture stress is the highest for sample G, followed by F, and least in D for this range of crack depth. This is because the magnitude of $|\sigma_{rcs\_G}| > |\sigma_{rcs\_F}| > |\sigma_{rcs\_D}|$. Also in Figure 13, if we consider a fixed crack of depth c = 40μm within the deep crack domain ( c > 20μm) and compare specimen A which has a high CS with specimen G which has a much deeper DOL but lower surface CS than A, the fractures stress of A is found to be significantly lower than that of G. This is because for the flaw depth c = 40μm, the magnitude of $|\sigma_{rcs\_A}| \ll |\sigma_{rcs\_D}|$. It is to be noted that the fracture stress corresponding to c = 40μm for specimen G is obtained by extrapolating the trendline in Figure 13 to capture the 40μm depth, since the crack depth generated by the 15psi abrasion pressure is below 40μm for specimen G; higher abrasion pressures might be needed to create a 40μm crack depth.

Therefore, from the observations in Figure 13, it can be implied that a high CS results in the highest fractures stress in the presence of shallow cracks and high DOL showed the highest fracture resistance in the presence of deeper cracks. Thus, the results reflect the importance of a high surface CS (with just enough DOL to contain the crack) for improving fracture resistance in the presence of shallow cracks, and the importance of a deep DOL in the situation of deeper cracks generating events. Hence, a reasonable combination of high surface CS and deep DOL is expected to provide the ultimate benefit in terms of mitigating fracture from both shallow and deeper cracks in the chemically strengthened glass.

4.4. Apparent fracture toughness, $K_{IC}^a$, as a function of surface CS at constant DOL

To understand the effect of surface CS on toughening of a chemically strengthened glass (at constant DOL) for different crack depths, the relation between $K_{IC}^a$ versus surface CS was plotted for the chemically strengthened specimens A-E in Figure 14. The samples all have an approximately constant DOL (35 ± 3μm) with varying surface CS. Two different crack size ranges are considered. Shallow crack (c < 20μm) created by the 1psi abrasion pressure, and deep crack (c > 20μm) created by the 15psi abrasion pressures. The data was also compared with that of the non-chemically strengthened glass, H (CS = 0MPa, DOL = 0μm).

For the 1psi abrasion having a shallow crack depth (recall Figure 11a) well contained within the DOL, we observe that the apparent fracture toughness $K_{IC}^a$ increases as the surface CS increases. The specimen A, with the highest CS, is observed to have the highest $K_{IC}^a$ and least for specimen E (See Figure 14). This is because for the crack depth range, the magnitude of the integral residual compressive stresses between crack tip and the surface are highest for specimen A and the least for specimen E as previously discussed in section 4.3. On the other hand, for the 15psi abrasion pressure which creates a deeper crack depth, the $K_{IC}^a$ for the specimens is observed to decrease significantly relative to the 1psi abrasion, and remain approximately constant regardless of the increasing trend in the surface CS. This is because at this deeper crack depth domain, the contribution of the compressive residual stress is significantly reduced, as the crack tip lies outside the DOL. This shows that in a damage event where shallow crack depth is introduced, the criterion for optimal performance in terms of fracture toughness is more a function of high surface CS than a deep DOL.

Furthermore, it is also shown in Figure 14 that while $K_{IC}^a$ for the chemically strengthened specimens decrease, as the crack transitions from shallow crack (1psi abrasion) to deeper crack (15psi abrasion) domain, $K_{IC}^a$ of the non-chemically strengthened glass remained approximately constant, with an average value of $0.64 \pm 0.08\ MPa\text{-}m^{0.5}$. Since there is no residual stress in a non-chemically strengthened glass, the apparent toughness is independent of the crack depth as expected. The consistency of the measured fracture toughness of the non- strengthened glass when compared to the previously reported value of $0.67\ MPa\text{-}m^{0.5}$ for this composition using the chevron notch short-bar technique [33], validates the accuracy of the experimental method (abrasion plus ring-on-ring technique) used in this work.

## 4.5 Apparent fracture toughness, $K_{IC}^a$, as a function of DOL per crack depth at constant surface CS

To understand the effect of varying DOL on toughening of a chemically strengthened glass (at constant surface CS) for different crack depth, the relation between $K_{IC}^a$ versus DOL per crack depth was plotted for chemically strengthened samples D, F and G in Figure 15. An approximately constant surface CS of $(463 \pm 4\ MPa)$ was considered with two ranges of crack depth. Shallow and deep crack depth domain created by the 1psi and 15psi abrasion pressures respectively, as previously mentioned in section 4.4 (recall Figure 11b). This also included the non-chemically strengthened glass for comparison.

Figure 15 shows that at a constant surface CS, increasing DOL increases the toughness of the glass for the 1psi created crack depths, as demonstrated by the chemically strengthened specimens D, F, and G. This is because the crack depth is still well within the DOL and the applied tensile stress is needed to overcome the residual compressive stress to further propagate the crack. As the abrasion pressure goes to 15psi and crack depth goes deeper, the apparent fracture toughness $K_{IC}^a$ of the specimen D decreases significantly compared to F and G. This is because the crack depth for specimen F and G are still well within the DOL, but have exceeded the DOL for D. This shows that in a damage event creating deep cracks, the criterion for better performance in terms of resistance to fracture is weighted more towards having a deep DOL than high surface CS.

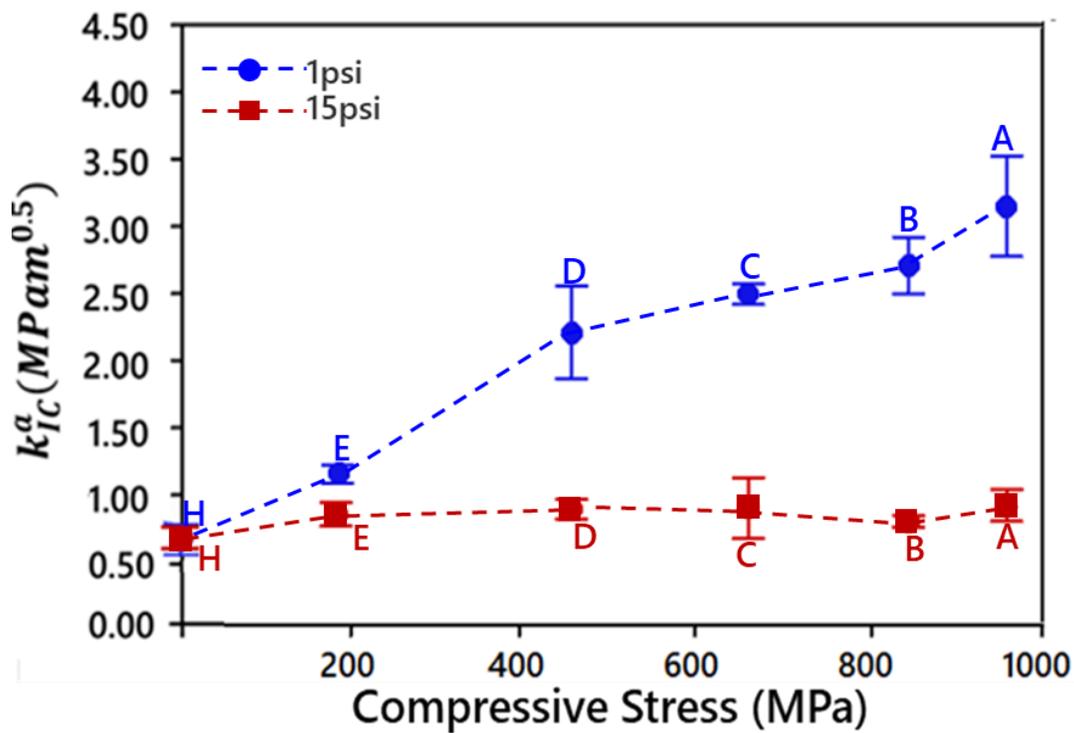

**Figure 14:** $K_{IC}^a$ as a function of surface CS for 1psi and 15psi abrasion pressure damage at constant DOL = $35 \pm 3\mu m$.

Another key observation with regard to Figures 14 and 15, is that there is a significant change in $K_{IC}^a$ for the chemically strengthened glass compared to the non-strengthned glass. This change is reflected due to the combined effect of the steepness in residual stress profile created by the ion-exchange process (CS-DOL relationship), and the severity of the crack size relative to the DOL. For specimen having a steep residual compressive stress profile ( specimen A-D), a small change in the crack depth ( as the crack transition from shallow to deep crack domain) results into a significant change in the magnitude of $K_{IC}^a$, while for a less steep residual compressive stress profile (specimen F and G), the change in $K_{IC}^a$ is minimal or negliblble.

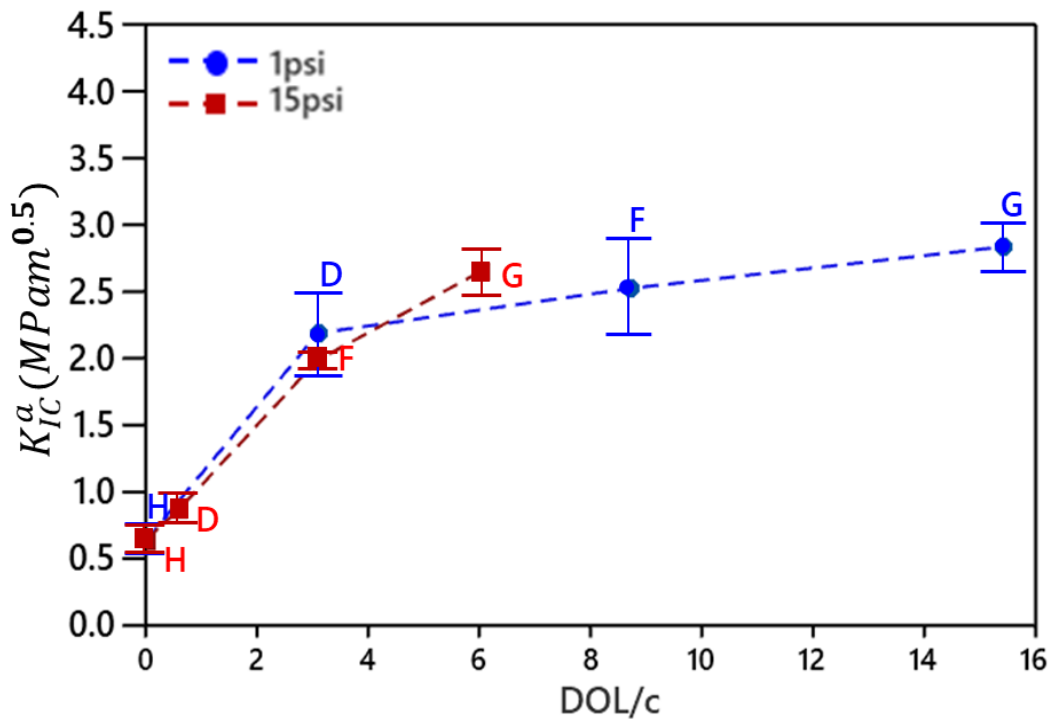

**Figure 15:** $K_{IC}^a$ *as a function of DOL per crack depth for 1psi and 15psi abrasion pressure damage at constant CS = 463 ± 4MPa.*

4.6. Apparent fracture toughness, $K_{IC}^a$ , as a function of CT

In previous section 4.4 and 4.5 it was shown that the prevention of failures from shallow and deep flaws are favored by the presence of high surface CS and deep DOL respectively. Central tension,

CT, on the inside of the glass, which represent the stored energy per unit volume in the glass due to the combine effect of surface CS and DOL for a given specimen thickness as shown in by the relation in eqn. (1), is evaluated as a function of the apparent toughness in this section. The $K_{IC}^a$ of the specimens A-H is plotted versus the CT as shown in Figure 16. It is observed that $K_{IC}^a$ increases proportionally with increase in CT for the 1psi abrasion (shallow crack) damage. However, in the presence of a deep crack (15psi abrasion damage) no net increase of statistical significance is observed between 0 to 45 MPa CT range. This is the because at this range the 1psi crack depth had likely exceeded the DOL of the specimen. However, as the CT increases beyond 60MPa, $K_{IC}^a$ is observed to also increase significantly. This is because the DOL corresponding to this CT is significantly greater than the 15psi created crack depth. These findings in addition to previously discussed in sections 4.4 and 4.5 suggests that in order to mitigate failure resulting from shallow flaws high CS, low CT or DOL enough to contain the crack is preferred than having a deep DOL. Deep DOL or high CT at surface CS at least 50% maximum surface CS is preferred in designing against failures from deep flaws.

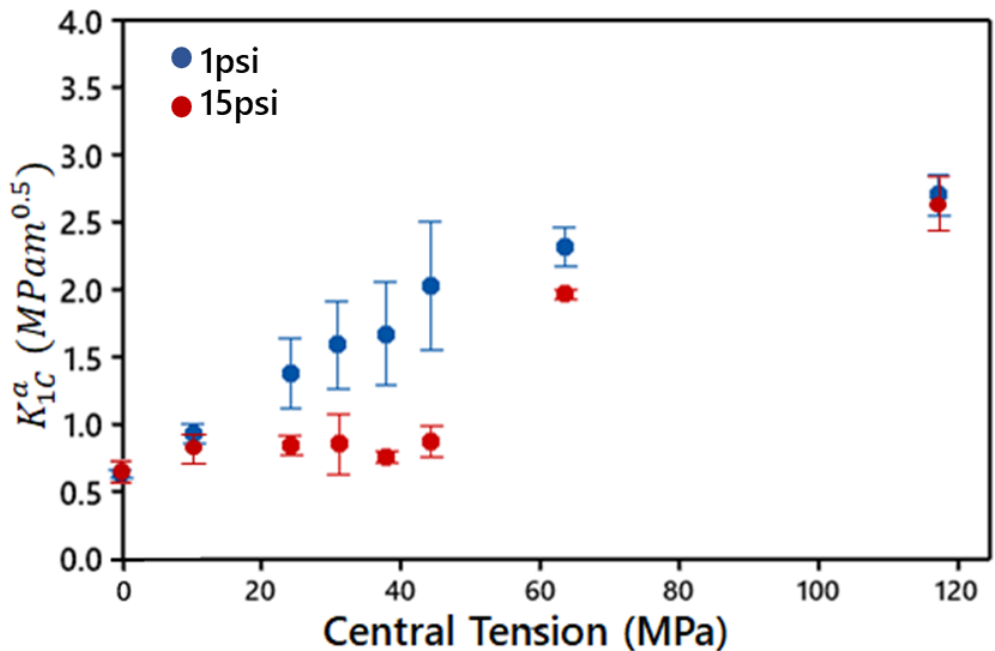

*Figure 16:* $K_{IC}^a$ *as a function of CT for 1psi and 15psi abrasion pressure damage*

## 4.7 Experimental results versus analytical model predictions of fracture toughness for shallow and deep cracks

The average apparent fractures toughness determined from experiments was compared with the analytically computed values obtained using Green's function-based method (computed using the eqn. (13) and (14) and the crack tip residual stress approach. The comparison of the $K_{IC}^a$ values was done as a function of the surface CS (keeping the DOL constant), and as a function of DOL (keeping the surface CS as constant), for two different crack depths as 10μm and 40μm respectively. It should be noted that the experiment $K_{IC}^a$ at 10μm and 40μm was computed by considering the fracture stress corresponding to these crack depths (Figure 13).

Figures 17a,b show the comparison between $K_{IC}^a$ obtained from experiments, Green's function-based prediction, and that obtained using the crack tip residual stress approach, as a function of surface CS at a constant DOL, for the two different crack depths, respectively. In Figure 17a, it is observed that for a crack depth c = 10μm, within the shallow crack domain, the experimentally determined $K_{IC}^a$ shows similar trend as the analytically computed values. All the three methods indicated that for a shallow crack (c =10μm), $K_{IC}^a$ increases with increase in surface CS for a fixed DOL. It is also observed that $K_{IC}^a$ computed using the crack tip residual stress approach shows better agreement with the experimental results compared to the Green function's approach. A plausible reason for this, could be the assumption of a linear residual stress profile in the Green's function approach, as opposed to the actual non-linear nature of the residual compressive stress profile while computing $K_r$. For the crack depth c = 40μm within the deeper crack domain, $K_{IC}^a$ is observed to increase less significantly with increase in surface CS (see Figure 17b). Similar trend is shown by the experimental results and the Green's function-based prediction, but the deviation between the two is higher compared to the case of c =10μm. The deviation between the experiment and the Green's

function-based computation can be further improved by considering the actual nonlinear residual compressive stress profile in a finite element-based simulation (considered to be a future work). The crack tip residual stress approach in this case is observed to yield a decreasing $K_{IC}^a$ value for an increasing CS. This is due to the fact that the crack tip residual stress method accounts only for the compressive stress contribution at the crack tip, as opposed to the integral of the compressive stress along the entire crack depth. Hence, for the case when the crack length exceeds the DOL, the stress at the crack tip ceases to be compressive in nature and becomes tensile, consequently reducing the resultant fracture toughness as seen in Figure 17b.

Figures 18 shows the comparison of the $K_{IC}^a$ obtained from the three different methods, as a function of surface DOL at a constant surface CS, in the event of a shallow crack damage of depth c=10μm. It is observed that all the three methods have shown similar trends in $K_{IC}^a$ values as a function of increasing DOL. The crack tip residual stress approach shows a much better agreement with experiments compared to the Green's function approach. This is likely due to the linear residual compressive stress profile distribution assumed in the Green's function $K_r$ computation. It is important to note that the comparison of experimentally measured $K_{IC}^a$ as a function of DOL (at constant surface CS), versus the analytical approaches, at a crack depth c = 40μm has not been done, as the crack depths generated by the 15psi abrasion pressure were below 40μm for the deeper DOL samples F and G. Higher abrasion pressures might be needed to achieve 40μm crack depth, which could be a topic for future study.

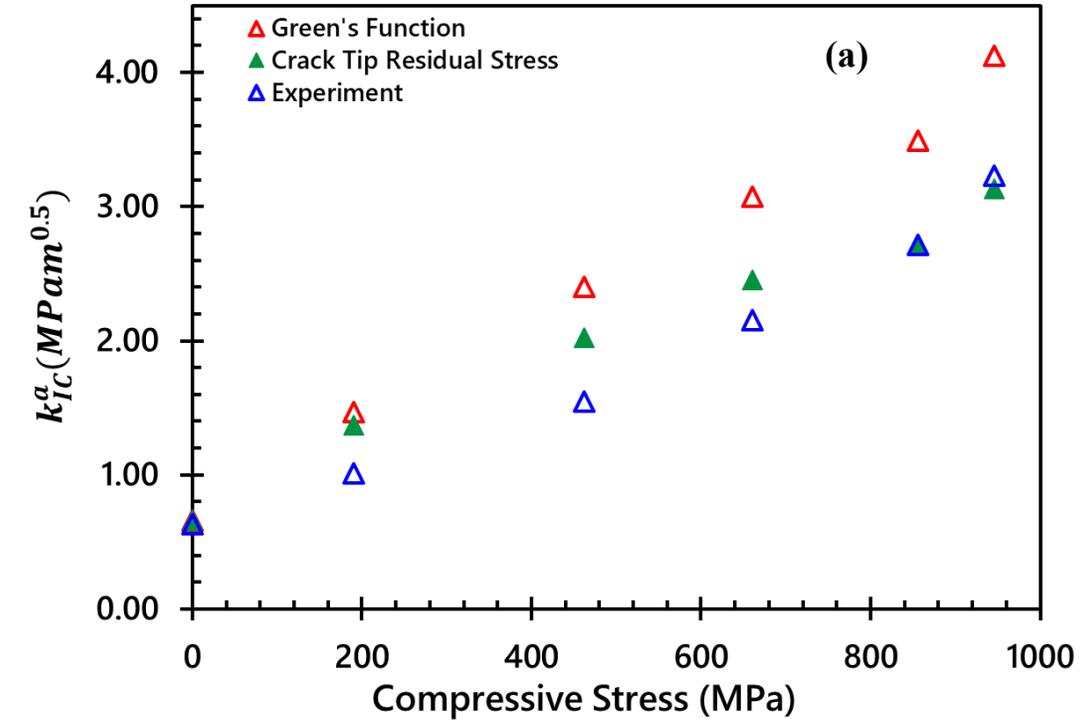
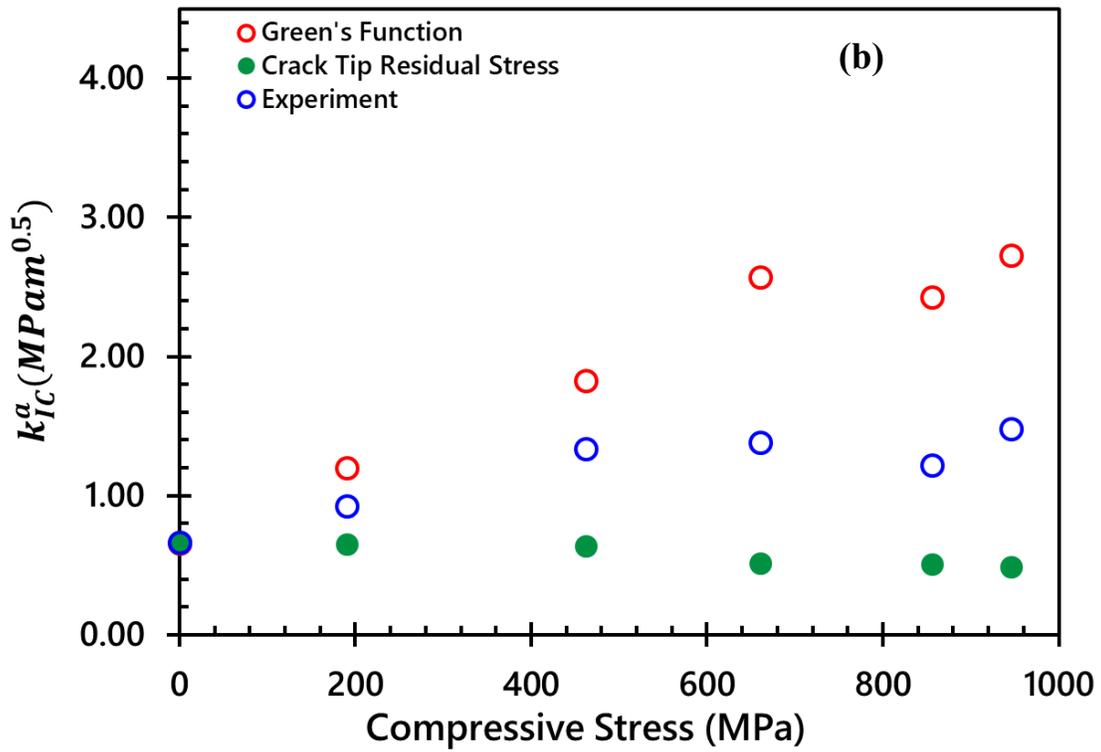

*Figure 17:* Comparison of experimental vs analytical $K_{IC}^a$ as a function of surface CS at constant DOL (35 ± 3μm) for (a) crack depth, c = 10μn (b) crack depth, c = 40μm.

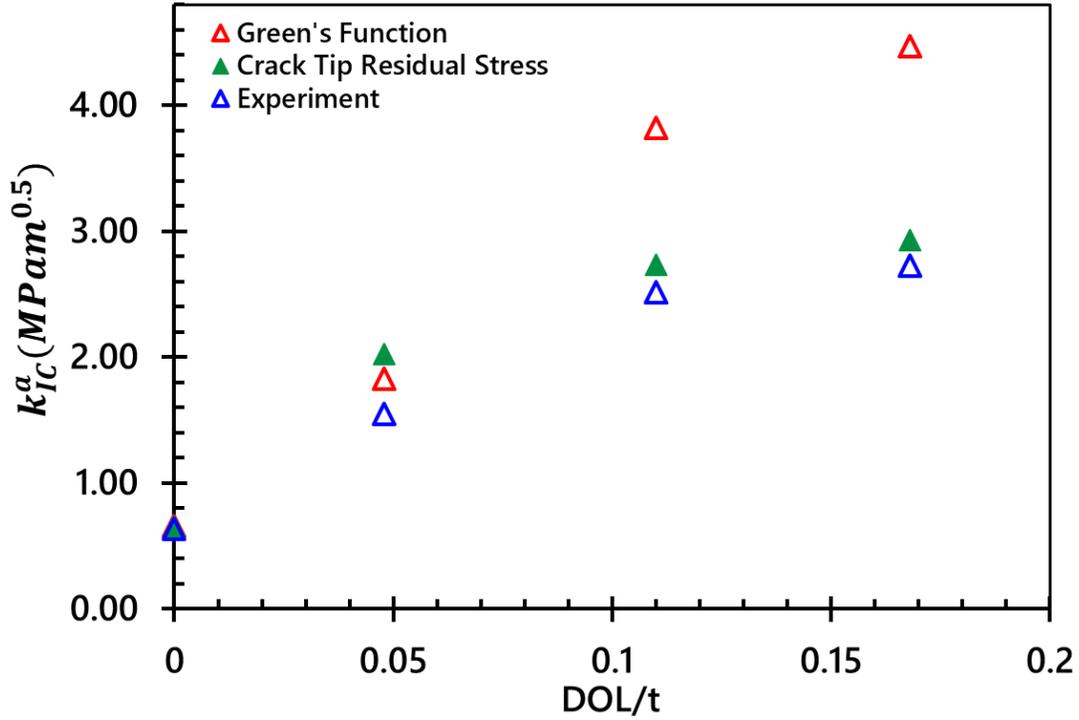

*Figure 18: Comparison of experimental vs analytical $K_{IC}^a$ as a function of DOL at constant Surface CS (460± 5MPa) for a given crack depth of c = 10μm.*

## 4. Conclusion

In the present work, experimental fracture stress and fracture toughness are measured via ring-on-ring test for a non-chemically strengthened (as-received) and a chemically strengthened sodium aluminosilicate (SAS) glass, abraded at different abrasion pressures, covering a wide range of crack depth (shallow to deep). The tests are done for various combinations of CS-DOL-CT for the ion-exchanged glass. The effect of the CS-DOL-CT combination on the fracture toughness of chemically strengthened glass at different crack depths is investigated and the results are discussed thoroughly.

It is determined that the mean fracture toughness of the as-received SAS glass used in this study is $0.64 \pm 0.08$ $MPa\text{-}m^{0.5}$, which is consistent with the measurements from previously reported work, where other experimental techniques were used. The apparent toughness (resistance to crack growth) of chemically strengthened SAS glass is seen to be higher than that of the "as-received" non-chemically strengthened glass. It is determined that increasing surface CS at a constant DOL or increasing DOL at a constant surface CS leads to an increase in toughness. The increase in toughness is quite significant when the crack depth is well within the DOL of the ion-exchanged glass.

For a shallow crack, which may be generated on the glass due to processing, finishing, or handling, a high surface CS, low CT or DOL deeper than the crack depth, enough to contain the crack, provides the optimum resistance against fracture. On the other hand, for a deep crack common in more aggressive in-field application of the thin glass, such as in consumer electronics (tablets, laptops, smartphones etc.) where the glass may be subjected to a sharp contact damage such as impact on rough surfaces, a moderate CS (~50% maximum CS), high CT, and high DOL would provide the best performance.

Lastly, the fracture toughness measured from experiments is compared to the analytical model prediction. The correlation between the experimental fracture toughness and the model predictions show reasonably good agreement for shallow cracks than the deeper cracks.

In summary, this study demonstrates that SiC abrasion plus RoR testing may be used to experimentally measure the fracture toughness of glass. It also establishes the fact that to prevent catastrophic failure in SAS glass, in application, chemical strengthening is a fruitful technique to improve the toughness provided the right CS-DOL-CT combination is incorporated. The right combination essentially depends on the crack sizes being considered in the fracture events. Hence, the choice of CS-DOL-CT combination is not unique for a given chemically strengthened glass. It should be designed appropriately to obtain the required damage resistance and toughness attributes necessary for the intended application.

**Acknowledgement**

The Authors would like to thank Corning® Research and Development Corporation product performance and reliability group for providing access to their laboratory to conduct this experiment and for their technical support on the mechanical testing.

**References:**

[1] D. R. Uhlmann and N. Kreidl, Glass Science and Technology, vol. 5, New York: Academic Press, 1980, pp. 133 -26.

[2] O. van der Velde, "Finding the Strength of Glass: A mechanical and Fractographic Research of Glass Biaxial Strength for Structural Purpose," Delft University of Technology, Delft, 2015.

[3] A. K. Varshneya and J. C. Mauro, Fundamentals of Inorganic Glasses, Sheffield: Elsevier, 2019.

[4] A. A. Griffith, "The Phenomena of Reture and Flow in Solids," vol. 221, pp. 163 -198, 1921.

[5] R. Gy, "Ion exchange for glass strengthening," *Material Scvience and Engineering,* vol. 149, no. 2, pp. 159-165, 2008.

[6] A. K. Varshneya, "Chemical Strenthening of Glass: Lesson Learned and yet to be Learned," *International Journal of Applied Glass Science,* vol. 1, no. 2, pp. 131-142, 2010.

[7] S. Gomez, A. J. Ellison and M. J. Dejneka, "Designing Strong Glass for Mobile Devices," in *SID 09 DIGEST*, Boston, 2012.

[8] M. Mirkhalaf, A. K. Dastjerdi and F. Barthelat., "Overcoming the brittleness of glass through bio-inspiration and micro-architecture," *Nature Communications,* vol. 3166, no. 5, 2014.

[9] L. Jiang, Y. Wang, I. Mohagheghian, X. Li, X. Guoa, L. Li, J. P. Dear and Y. Yan, "Effect of residual stress on the fracture of chemically strengthened thin aluminosilicate glass," *Journal of Materials Science,* vol. 52, no. 3, p. 1405–1415, 2016.

[10] S. S. Kistler, "Stresses in Glass Produced by Nonuniform Exchange of Monovalent Ions," *Journal of the American Ceramic Society,* vol. 45, no. 2, pp. 59-68, 1962.

[11] . V. Leboeuf, J. P. Blondeau , D. De Sousa Meneses and O. Véron, "Potassium ionic exchange in glasses for mechanical property improvement," *Journal of Non-Crystalline Solids,* vol. 377, pp. 60 -65, 2012.

[12] D. K. Hale, "Strengthening of Silicate Glasses by Ion Exchange," *Nature,* vol. 217, no. 5134, pp. 1115-1118, 1968.

[13] S. Karlsson, "Modification of Float Glass Surface By Ion Exchange," Linnaeus University Press, Växjö, 2012.

[14] R. Dugnani, "Residual stress in ion-exchanged silicate glass: An analytical solution," *Journal of Non-Crystaline Solids,* vol. 471, pp. 368 - 378, 2017.

[15] A. Tandia, K. D. Vargheese and J. C. Mauro, "Elasticity of ion stuffing in chemically strengthened glass," *Journal of Non-Crystalline Solids,* vol. 358, pp. 1569 - 1574, 2012.

[16] X. Guo, A. L. Pivovarov, M. M. Smedskjaer, M. Potuzak and J. C. Mauro, "Non-conservation of the total alkali concentration in ion-exchanged glass," *Journal of Non-Crystalline Solids,* vol. 387, pp. 71-75, 2014.

[17] A. I. Fu and J. C. Mauro, "Mutual diffusivity, network dilation, and salt bath poisoning effects in ion-exchanged glass," *Journal of Non-Crystalline Solids,* vol. 363, pp. 199 - 204, 2013.

[18] C. Calahoo, J. W. Zwanziger and I. S. Butler, "Mechanical−Structural Investigation of Ion-Exchanged Lithium Silicate Glass using Micro-Raman Spectroscopy," *Journal of Physical Chemistry,* vol. 120, p. 7213−7232, 2016.

[19] H. Sun and R. Dugnani, "A study on ion-exchanged, soda-lime glass's residual stress relationship with K+/Na+ concentration," *International Journal of Applied Glass Science,* vol. 11, no. 1, pp. 134 - 146, 2020.

[20] J. Shen, D. J. Green, R. E. Tressler and D. L. Shelleman, "Stress Relaxation of a Soda Lime Silicate Glass Below the Glass Transition Temperature," *Journal of Non-Crystalline Solids,* vol. 324, no. 4, pp. 277 -288, 2003.

[21] C. Chao, J. Zonghe and C. Jingyi, "Influence of Surface Structure on the Fracture Toughness of Tempered Glass," *Journal of American Ceramic Society,* vol. 71, no. 10, pp. 434 -435, 1988.

[22] G. Macrelli, "Chemically strengthened glass by ion exchange: Strength evaluation," *International Journal of Applied Glass Science,* vol. 9, no. 2, pp. 156-166, 2018.

[23] D. Connolly, "Fracture Analysis of Chemically Strengthened Glass Disks," *Journal of the American Ceramic Society,* vol. 72, no. 7, pp. 1162-1166, 1989.

[24] F. Petit, A. Sartieaux, M. Gonon and F. Cambier, "Fracture toughness and residual stress measurements in tempered glass by Hertzian indentation," *Acta Materialia*, vol. 55, no. 8, pp. 2765-2774, 2007.

[25] I. Erdem, D. Guldiren and S. Aydin, "Chemical tempering of soda lime silicate glasses by ion exchange process for the improvement of surface and bulk mechanical strength," *Journal of Non-Crystalline Solids,* vol. 473, pp. 170-178, 2017.

[26] D. J. Green and B. R. Maloney, "Influence of Surface Stress on Indentation Cracking," *Journal of the American Ceramic Society,* vol. 69, no. 3, pp. 223 - 225, 1986.

[27] E. M. Aaldenberg, P. J. Lezzi, J. H. Seaman, T. A. Blanchet and M. Tomozawa, "Ion-Exchanged Lithium Aluminosilicate Glass: Strength and Dynamic Fatigue," *Journal of the American Ceramic Society,* vol. 99, no. 8, pp. 2645-2654, 2016.

[28] L. Jiang, Y. Wang, I. Mohagheghian, X. Li, X. Guoa and L. Li, "Subcritical crack growth and lifetime prediction of chemically strengthened aluminosilicate glass," *Materials and Design,* vol. 122, pp. 128 - 135, 2017.

[29] H. Morozumi, "Crack Initiation Tendency of Chemically Strengthened," *International Journal of Applied Glass Science,* vol. 6, no. 1, pp. 64-71, 2015.

[30] K. Hayashi, S. Akiba, T. Sakagami, S. Sawamura, T. Nakashima and H. Ohkawa, "Cover Glass for Mobile Devices," in *Society for Information Display*, Boston, 2012.

[31] H. Ikeda, K. Kinoshita, M. Fukada, K. Kawamoto, T. Murata, K. Choju, M. Ohji and . H. Yamazaki, "High Edge Strength Glass for Mobile Devices," in *Society for Information Display*, San Jose, 2015.

[32] J. Amin, B. O. Egboiyi, J. D. Pesansky, K. B. Reiman, R. V. Roussev and B. P. Strines, "Strengthened Glass With Deep Depth of Compression". United States Patent 10150698, 11 December 2018.

[33] "Corning® Gorilla® Glass 4 Product Information Sheet," Corning Incorporated , September 2015. [Online]. Available: https://www.corning.com/microsites/csm/gorillaglass/PI_Sheets/CGG_PI_Sheet_Gorilla%20Glass%204.pdf. [Accessed 4 March 2020].

[34] D. C. Allan, K. W. KochIII, R. V. Roussev, R. A. Schaut and V. M. Schneider, "Systems and methods for measuring the stress profile of ion-exchanged glass". United States of America Patent 9140543B1, 22 September 2015.

[35] IEC61747-40-6, "Liquid crystal display devices - Part 40-6: Mechanical testing of display cover glass for mobile devices - Retained biaxial flexural strength (abraded ring-on-ring)," IEC, Worcester, 2018.

[36] G. Quinn, Fractography of Ceramics and Glasses, Gaithersburg: NIST, 2016.

[37] ASTM C1499-09, "Standard test method for Monotonic Equilibrium Flexural Strength of Advanced Ceramic at Ambient Temperature," ASTM International, West Conshohocken, 2009.

[38] S. Timoshenko and S. Woinowsky-Krieger, Theory of Plates and Shells, 2nd, Ed., New York: McGraw Hill, 1959.

[39] T. L. Anderson, Fracture Mechanics Fundermentals and Application, FL: CRC Press, 2005.

[40] ASTM C1322-05b, "Standard practice for Fractography and Characterization of Fracture Origins in advanced Ceramics," ASTM, West Conshohocken, 2005.

[41] V. M. Sglavo and L. Larentis, "Flaw-Insensitive Ion-Exchanged Glass: I, Theoretical Aspects," *Journal of the American Ceramic Society,* vol. 84, no. 8, pp. 1827-1831, 2001.

[42] R. F. Cook and D. R. Clarke, "Fracture Stability, R-Curves and Strength Variability," *Acta Metallurgica,* vol. 36, no. 3, pp. 555-562, 1988.

[43] D. J. Green, An Introduction to the Mechanical Properties of Ceramics, Cambridge, UK: University Press, 1998.

[44] L. Ma and R. Dugnani, "Fractographic analysis of silicate glasses by computer vision," *Journal of the European Ceramic Society,* vol. 40, no. 8, pp. 3291-3303, 2020.

[45] K. Kanehara, K. Imakita, Y. Kobayashi and A. Koike, "Analysis of Fracture Mechanism in Sand Paper Drop Test for Cover Glass," in *Society for Information Display*, San Jose, 2019.